\def\Title{Local Transition Functions of Qunatum Turing Machines}
\def\Author{Masanao Ozawa and Harumichi Nishumura}

\documentstyle[12pt,mystyle]{article}

  \newcommand{\beq}{\begin{equation}}
  \newcommand{\eeq}{\end{equation}}
  \newcommand{\beql}[1]{\begin{equation}\label{eq:#1}}

  \newcommand{\beqa}{\begin{eqnarray}}
  \newcommand{\eeqa}{\end{eqnarray}}
  \newcommand{\beqas}{\begin{eqnarray*}}
  \newcommand{\eeqas}{\end{eqnarray*}}

  \newtheorem{Theorem}{Theorem}[section]
  \newtheorem{Proposition}[Theorem]{Proposition}
  \newtheorem{Lemma}[Theorem]{Lemma}

  \newenvironment{Proof}{\begin{trivlist}
    \item[\hskip \labelsep {\em \indent Proof.}]}{\qed\end{trivlist}}
  \newcommand{\qed}{{\em QED}}

  \newcommand{\C}{{\bf C}}

  \newcommand{\Z}{{\bf Z}}

  \newcommand{\beqan}{\begin{eqnarray*}}
  \newcommand{\beqar}[1]{\begin{equation}\label{#1}\begin{array}{l}}

  \newcommand{\eeqar}{\end{array}\end{equation}}

\newcommand{\bra}[1]{\langle#1|}
\newcommand{\ket}[1]{|#1\rangle}
\newcommand{\braket}[1]{\langle#1\rangle}

  \title{{\bf \Title}\thanks{Main results of this work were presented at 
the 4th International Conference on Quantum Communication, Computing, 
and Measurement (Evanston, IL, August 22-27, 1998) by the first author 
and appeared in {\em Quantum Communication, Computing, and Measurement 2}, 
edited by P.\ Kumar et al., Plenum, New York, 2000, pp.~241--248.}
}
  \author{\sc \Author \\ 
  \small\em  Graduate School of Human Informatics
and School of Informatics and Sciences \\
  \small\em  Nagoya University, Chikusa-ku, Nagoya 464-8601, Japan\\
and \\
CREST, Japan Science and Technology}
  \date{}
\begin{document}
\maketitle
\begin{abstract}
Foundations of the notion of quantum Turing machines are
investigated. According to Deutsch's formulation, the time evolution of 
a quantum Turing machine is to be determined by the local
transition function.  In this paper, the local transition 
functions are characterized for fully general quantum Turing 
machines, including multi-tape quantum Turing machines, 
extending the results due to Bernstein and Vazirani.
\end{abstract}



\date{}

\section{Introduction} 
Feynman \cite{Fey82} pointed out that a Turing machine
cannot simulate a quantum mechanical process efficiently and 
suggested that a computing machine 
based on quantum mechanics might be more powerful than Turing machines.
Deutsch introduced quantum Turing machines \cite{Deu85} and 
quantum circuits \cite{Deu89} for establishing the notion
of quantum algorithm exploiting ``quantum parallelism''.
A different approach to quantum Turing machines was taken
earlier by Benioff \cite{Beni80} based on the Hamiltonian description 
of Turing machines.
Bernstein and Vazirani \cite{BV97} instituted quantum complexity 
theory based on quantum Turing machines and constructed an
efficient universal quantum Turing machine.
Yao \cite{Yao93} reformulated the quantum circuit models
 by singling out the acyclic ones and  
showed that a computation by a quantum Turing 
machine can be simulated by a polynomial size quantum circuit.
The search for an efficient quantum algorithm for a well-studied 
but presumably intractable problem was achieved strikingly by 
Shor \cite{Sho94}, who found bounded error probability quantum 
polynomial time algorithms for the factoring problem 
and the discrete logarithm problem.

In this paper, foundations of the concept of quantum Turing machines
are examined.
In Deutsch's formulation \cite{Deu85},
a quantum Turing machine is defined to be a quantum system 
consisting of a processor, a moving head, and a tape, obeying a unitary 
time evolution determined by local interactions between its components. 
The machine is then allowed to be in a superposition of 
computational configurations. 
Deutsch \cite{Deu85} pointed out that the global transition function
between computational configurations should be determined by a local 
transition function which depends only on local configurations. 
Bernstein and Vazirani \cite{BV97} found a simple 
characterization of the local transition functions for the restricted
class of quantum Turing machines in which the head must move either 
to the right or to the left at each step.  
Since the above characterization constitutes an alternative
definition of quantum Turing machines more tractable in the field of
theoretical computer science, it is an interesting problem to find
a general characterization valid even when the head is not required 
to move or more generally when the machine has more than one tape.
The purpose of this paper is to solve this problem, 
while for this and foundational purposes we also provide a completely 
formal treatment of the theory of quantum Turing machines. 
Extending the Bernstein-Vazirani theory \cite{BV97}, 
the computational complexity theory for general quantum Turing machines 
defined by the conditions given in this paper 
will be published in our forthcoming paper \cite{NO99}. 

The paper is organized as follows.
In Section 2, quantum Turing machines are introduced along with
Deutsch's original formulation.
We extend Deutsch's formulation to the case where the head is not 
required to move every step.   
In Section 3, the local transition functions of quantum Turing 
machines are introduced along with Deutsch's requirement of
operations by finite means and the problem of the characterization 
of local transition functions is formulated. 
In Section 4, quantum Turing machines are formulated as mathematical
structures and we prove a characterization theorem of the local
transition functions of quantum Turing machines.
We adopt here the column vector approach, where the characterization
is obtained from the requirement that the column vectors of the 
transition matrix are orthonormal.
 In Section 5, we prove an alternative characterization theorem of 
the local transition functions along with the row vector approach.
In Section 6, the characterization is extended to multi-tape 
quantum Turing machines.

\section{Quantum Turing machine as a physical system}

A {\em quantum Turing machine} ${\cal Q}$ is a quantum system consisting of 
a {\em processor}, a bilateral infinite {\em tape}, 
and a {\em head} to read and write a symbol on the tape.  
Its configuration is determined by the {\em processor configuration} $q$ 
from a finite set $Q$ of symbols,
the {\em tape configuration} $T$ represented by an infinite string from a
finite set $\Sigma$ of symbols,
and the discretized {\em head position} $\xi$ taking values in the set $\Z$ 
of integers.
The tape consists of {\em cells} numbered by the integers.
The head position $\xi\in\Z$ stands for the place of the cell numbered 
by $\xi$. 
We assume that 
$\Sigma$ contains the symbol $B$ representing 
the blank cell in the tape.
For any integer $m$ the symbol at the cell $m$ on 
the tape is denoted by $T(m)$.
We assume that the possible tape configurations are such that
$T(m)=B$ except for finitely many cells $m$. 
The set of all the possible tape configurations is denoted by
$\Sigma^{\#}$.  The set $\Sigma^{\#}$ is a countable set.
Thus, any configuration $C$ of ${\cal Q}$ is represented by a triple
$C=(q,T,\xi)$ in the configuration space 
${\cal C}(Q,\Sigma)=Q\times\Sigma^{\#}\times\Z$.
The quantum state of ${\cal Q}$ is represented by 
a unit vector in the Hilbert 
space ${\cal H}(Q,\Sigma)$ generated by the configuration space 
${\cal C}(Q,\Sigma)$ so that the vectors in ${\cal H}(Q,\Sigma)$ 
can be identified with the square summable complex-valued functions 
defined on $Q\times\Sigma^{\#}\times\Z$. 
The complete orthonormal basis canonically in one-to-one correspondence 
with the configuration space is called the {\em computational basis}.
Thus, the computational basis is represented by 
$\ket{C}=\ket{q,T,\xi}$
for any configuration $C=(q,T,\xi)\in {\cal C}(Q,\Sigma)$.

In classical physics, physical quantities are represented by   
real-valued functions defined on the phase space 
coordinated by the configuration and the generalized momentum. 
In quantum mechanics, they are called observables and 
represented by self-adjoint operators 
on the Hilbert space of quantum states.  
The procedure to define the observables from the classical 
description of the system is usually called the quantization.
In order to define the observables quantizing the configurations,
we assume the numbering of the sets $Q$ and $\Sigma$ such that
$Q=\{q_{0},\ldots,q_{|Q|-1}\}$ and 
$\Sigma=\{\sigma_{0},\ldots,\sigma_{|\Sigma|-1}\}$,
where we denote by $|X|$ the number of the elements of a set $X$. 
We define observables $\hat{q}$, $\hat{T}(m)$ for $m\in\Z$, 
and $\hat{\xi}$ representing the processor configuration, 
the symbol at the cell $m$, and the head position, respectively, 
 as follows.
$$
\hat{q}=\sum_{n=0}^{|Q|-1}n\ket{q_{n}}\bra{q_{n}},\quad
\hat{T}(m)=\sum_{n=0}^{|\Sigma|-1}n\ket{\sigma_{n}}\bra{\sigma_{n}},\quad
\hat{\xi}=\sum_{\xi\in\Z}\xi\ket{\xi}\bra{\xi}.
$$

The computation begins at $t=0$ and
proceeds in steps of a fixed unit duration $\tau$. 
The dynamics of ${\cal Q}$ are described by a unitary operator 
$U$ on ${\cal H}(Q,\Sigma)$ which specifies 
the evolution of the system during a single {\em computational step} 
so that we have
$$
U^{\dagger}U=UU^{\dagger}=I,\quad\ket{\psi(n\tau)}=U^{n}\ket{\psi(0)}
$$
for all positive integers $n$.  

\section{Local transition functions}
Deutsch \cite{Deu85} required that the quantum Turing machine operates
finitely, i.e., (i) only a finite system is in motion during any one
step, (ii) the motion depends only on the quantum state of a local subsystem,
and (iii) the rule that specifies  the motion  can be given finitely in
the  mathematical sense.
To satisfy the above requirements, the matrix elements of $U$ are
required to take the following form\footnotemark:
\footnotetext{This condition is a natural extension of Deutsch's
condition \cite{Deu85} to the case where the head is not
required to move.
}
\begin{equation}\label{eq:31}
\bra{q',T',\xi'}U\ket{q,T,\xi}=\left\{
\begin{array}{ll}
\delta(q,T(\xi),q',T'(\xi),1)& \mbox{if}\ \xi'=\xi+1\\
\delta(q,T(\xi),q',T'(\xi),0)& \mbox{if}\ \xi'=\xi\\
\delta(q,T(\xi),q',T'(\xi),-1)& \mbox{if}\ \xi'=\xi-1
\end{array}
\right.
\end{equation}
whenever $T'(m)=T(m)$ for all $m\neq \xi$, 
and $\bra{q',T',\xi'}U\ket{q,T,\xi}=0$ otherwise, 
for any configurations $(q,T,\xi)$ and $(q',T',\xi')$.
The above condition ensures that the tape is changed 
only at the head position $\xi$ at the beginning of each computational step,
 and that during each step the head position cannot change
 by more than one unit. The function $\delta(q,T(\xi),q',T'(\xi),d)$, 
where $q,q'\in Q$, $T(\xi),T'(\xi)\in \Sigma$,
and $d\in\{-1,0,1\}$, represents a dynamical motion depending only on
the local observables $\hat{q}$ and $\hat{T}(\xi)$. 
It follows that the relation $\delta(q,\sigma,q',\tau,d)=c$ can be interpreted
as the following operation of ${\cal Q}$:
if the processor is in the configuration $q$ and 
if the head reads the symbol $\sigma$,
then it follows with the amplitude $c$ that
the processor configuration turns to $q'$,
the head writes the symbol $\tau$,
and that the head moves one cell to the right if $d=1$,
to the left if $d=-1$, or does not move if $d=0$. 
We call $\delta$ the {\em local transition function} 
of the quantum Turing machine ${\cal Q}$.

The local transition function $\delta$ can be arbitrarily given except 
for the requirement that $U$ be unitary.
Each choice defines a different quantum Turing machine ${\cal Q}[\delta]$ 
with the same configuration space ${\cal C}(Q,\Sigma)$.
Thus, if we have an intrinsic characterization of the local transition
function $\delta$, quantum Turing machines can be defined formally without 
referring to the unitary operator $U$ as a primitive notion.

From Eq.\ (\ref{eq:31}) the time evolution operator $U$ is determined
conversely from the local transition function $\delta$ by
\begin{equation}\label{eq:32}
U\ket{q,T,\xi}
=\sum_{p,\tau,d}\delta(q,T(\xi),p,\tau,d)
\ket{p,T^{\tau}_{\xi},\xi+d}
\end{equation}
for any configuration $(q,T,\xi)$, where $T^{\tau}_{\xi}$ is the tape
 configuration defined by 
$$
T^{\tau}_{\xi}(m)=\left\{
\begin{array}{ll}
\tau& \mbox{if}\ m=\xi,\\
T(m)& \mbox{if}\ m\not=\xi.
\end{array}
\right.
$$

Now we can formulate the characterization problem of local transition
functions of quantum Turing machines:
{\em Let $\delta$ be a complex-valued function on 
$Q\times\Sigma\times Q\times\Sigma\times\{-1,0,1\}$
and let $U$ be the operator on ${\cal H}(Q,\Sigma)$ defined by 
Eq.\ (\ref{eq:32}).
Then, what conditions ensure that the operator $U$ is unitary?}

This problem is answered by the following statement:
{\em The operator $U$ is unitary if and only if $\delta$ satisfies the 
following conditions.

{\rm (a)} For any $(q,\sigma)\in Q\times\Sigma$,
$$
\sum_{p,\tau,d}|\delta(q,\sigma,p,\tau,d)|^{2}=1.
$$

{\rm (b)} For any $(q,\sigma), (q',\sigma')\in Q\times\Sigma$
with $(q,\sigma)\ne (q',\sigma')$,
$$
\sum_{p,\tau,d}
\delta(q',\sigma',p,\tau,d)^{*}\delta(q,\sigma,p,\tau,d)=0.
$$

{\rm (c)} For any $(q,\sigma,\tau),(q',\sigma',\tau')
\in Q\times\Sigma^{2}$, we have
$$
\sum_{p\in Q,d=0,1}\delta(q',\sigma',p,\tau',d-1)^{*}
\delta(q,\sigma,p,\tau,d)=0.
$$

{\rm (d)} For any $(q,\sigma,\tau),(q',\sigma',\tau')
\in Q\times\Sigma^{2}$, we have
$$
\sum_{p\in Q}\delta(q',\sigma',p,\tau',-1)^{*}\delta(q,\sigma,p,\tau,1)=0.
$$
}
 
The proof will be given in the next section.
If it is assumed that the head must move either to the right or to 
the left at each step (two-way quantum Turing machines),
 the condition (c) is automatically satisfied.
In this case, the above statement is reduced to the result due to 
Bernstein and Vazirani \cite{BV97}.
In Section 5, we will also characterize the local transition functions 
of multi-tape quantum Turing machines.

In order to maintain the Church-Turing thesis, we need to require
that the unitary operator $U$ is constructive, or that the 
range of the local transition function $\delta$ is in the computable
complex numbers.
From the complexity theoretical point of view, we need also to
require that the matrix elements of $U$ are polynomially computable 
complex numbers, or that the range of the transition function $\delta$ is 
in the polynomially computable complex numbers.

\section{Quantum Turing machine as a mathematical structure}

In order to formulate the notion of a quantum Turing machine as a 
formal mathematical structure rather than a well-described
physical system, we shall introduce the following
mathematical definitions.
A {\em Turing frame\/} is a pair $(Q,\Sigma)$ of a finite set $Q$ 
and a finite set $\Sigma$ with a specific element denoted by $B$.
In what follows, let $(Q,\Sigma)$ be a Turing frame.
Let $\Sigma^{\#}$ be the set of functions $T$ from the set $\Z$ of integers 
to $\Sigma$ such that $T(m)=B$ except for finitely many $m\in\Z$.
The {\em configuration space} of $(Q,\Sigma)$
is the product set ${\cal C}(Q,\Sigma)=Q\times\Sigma^{\#}\times\Z$.

For any $(p,\tau,d)\in Q\times\Sigma\times\{-1,0,1\}$, denote by 
${\cal C}(p,\tau,d)$ the set of configurations
 $(p,T,\xi)\in{\cal C}(Q,\Sigma)$ such that
$T(\xi-d)=\tau$.
Let $(p,\tau,d)\in Q\times\Sigma\times \{-1,0,1\}$. 
We define the transformation $\alpha(p,\tau,d)$ from ${\cal C}(Q,\Sigma)$ to
${\cal C}(p,\tau,d)$ by
\begin{equation}\label{eq:41}
\alpha(p,\tau,d)(q,T,\xi)=(p,T_{\xi}^{\tau},\xi+d)
\end{equation}
for all $(q,T,\xi)\in{\cal C}(Q,\Sigma)$.
It is easy to see that $\alpha(p,\tau,d)$ represents the operation such
that the processor configuration turns to $p$,
the head writes the symbol $\tau$, and then moves with $|d|$ step to the 
direction $d$.
We define the transformation $\beta(p,\tau,d)$ from ${\cal C}(Q,\Sigma)$
to ${\cal C}(p,\tau,0)$ by
\begin{equation}\label{eq:42}
\beta(p,\tau,d)(q,T,\xi)=(p,T^{\tau}_{\xi-d},\xi-d)
\end{equation}
for any $(q,T,\xi)\in{\cal C}(Q,\Sigma)$.
It is easy to see that $\beta(p,\tau,d)$ represents the operation such
that the processor configuration turns to $p$, 
the head moves with $|d|$ step to the direction $-d$
and then writes the symbol $\tau$.
The following proposition can be checked by straightforward
verifications.

\begin{Proposition}\label{th:622a}
{\rm (i)} Let $d\in \{-1,0,1\}$.
 If $(q,\sigma)\not=(q',\sigma')\in Q\times\Sigma$ then 
${\cal C}(q,\sigma,d)\cap{\cal C}(q',\sigma',d)=\emptyset$ and 
$$
{\cal C}(Q,\Sigma)=\bigcup_{(q,\sigma)\in Q\times\Sigma}{\cal C}(q,\sigma,d).
$$

{\rm (ii)} Let $(q,\sigma,p,\tau,d)\in Q\times\Sigma\times
 Q\times\Sigma\times\{-1,0,1\}$.
We have
$$
\beta(q,\sigma,d)\alpha(p,\tau,d)C=C
$$
for all $C\in{\cal C}(q,\sigma,0)$ and
$$
\alpha(p,\tau,d)\beta(q,\sigma,d)C'=C'
$$
for all $C'\in{\cal C}(p,\tau,d)$.

{\rm (iii)} The mapping $\alpha(p,\tau,d)$
 restricted to ${\cal C}(q,\sigma,0)$ 
has the inverse mapping $\beta(q,\sigma,d)$
 restricted to ${\cal C}(p,\tau,d)$, i.e.,
$$
{\cal C}(q,\sigma,0)
\begin{array}{c}
\stackrel{\alpha(p,\tau,d)}{\longrightarrow}\\
\stackrel{\beta(q,\sigma,d)}{\longleftarrow}
\end{array}
{\cal C}(p,\tau,d).
$$
\end{Proposition}

\sloppy
A configuration $(q,T,\xi)$ is said to {\em precede\/} a configuration
$(q',T',\xi')$, in symbols $(q,T,\xi)\prec (q',T',\xi')$, 
if $T'(m)=T(m)$ for all $m\not=\xi$ 
and $|\xi'-\xi|\le 1$.

\begin{Proposition}\label{th:825a}
For any $C,C'\in{\cal C}(Q,\Sigma)$, the following 
conditions are equivalent.

{\rm (i)} $C\prec C'$.

{\rm (ii)} There is some $(p,\tau,d)\in Q\times\Sigma\times\{-1,0,1\}$
 such that $C'=\alpha(p,\tau,d)C$.

{\rm (iii)} There is some $(q,\sigma,d)\in Q\times\Sigma\times\{-1,0,1\}$
 such that $C=\beta(q,\sigma,d)C'$.
\end{Proposition}
\begin{Proof}
Let $C=(q,T,\xi)$ and $C'=(q',T',\xi')$.

(i)$\Rightarrow$(ii): If (i) holds, 
we have $C'=\alpha(q',T'(\xi),\xi'-\xi)C$
so that (ii) holds.

(ii)$\Rightarrow$(iii): Suppose that (ii) holds.
Since $C\in{\cal C}(q,T(\xi),0)$, by Proposition \ref{th:622a} (ii)
we have
$$
\beta(q,T(\xi),d)C'=
\beta(q,T(\xi),d)\alpha(p,\tau,d)C=C.
$$

(iii)$\Rightarrow$(i): If (iii) holds, we have 
$C=(p,T'{}_{\xi'-d}^{\sigma},\xi'-d)$ and hence $\xi'-\xi=d$
and $T(m)=T'{}_{\xi'-d}^{\sigma}(m)=T'(m)$ for $m\not=\xi'-d=\xi$
so that (i) holds.
\end{Proof}

The {\em quantum state space} of the Turing frame $(Q,\Sigma)$ is 
the Hilbert space ${\cal H}(Q,\Sigma)$ spanned by ${\cal C}(Q,\Sigma)$ with
 the canonical basis $\{\ket{C}|\ C\in{\cal C}(Q,\Sigma)\}$ called the
 {\em computational basis}.
 A {\em local transition function} for
 $(Q,\Sigma)$ is a function from
 $Q\times\Sigma\times Q\times\Sigma\times \{-1,0,1\}$
 into the complex number field $\C$.

In what follows, let $\delta$ be a local transition function for $(Q,\Sigma)$.
The {\em evolution operator} of $\delta$ is a linear operator $M_{\delta}$ on 
${\cal H}(Q,\Sigma)$ such that
\begin{equation}\label{eq:43}
M_{\delta}\ket{q,T,\xi}=
\sum_{p,\tau,d}\delta(q,T(\xi),p,\tau,d)\ket{p,T^{\tau}_{\xi},\xi+d}
\end{equation}
for all $(q,T,\xi)\in{\cal C}(Q,\Sigma)$; the summation $\sum_{p,\tau,d}$
is taken over all $(p,\tau,d)\in Q\times\Sigma\times \{-1,0,1\}$ above and
in the rest of this section unless stated otherwise. 
Eq.\ (\ref{eq:43}) uniquely defines the bounded operator $M_{\delta}$ 
on the space ${\cal H}(Q,\Sigma)$ as shown in Appendix A.

Let $(q,T,\xi),\ (q',T',\xi')\in{\cal C}(Q,\Sigma)$.
The following formula can be verified from Eq.\ (\ref{eq:43}) 
by straightforward calculation.

\begin{equation}\label{eq:44}
\bra{q',T',\xi'}M_{\delta}\ket{q,T,\xi}
=
\left\{
\begin{array}{ll}
\delta(q,T(\xi),q',T'(\xi),\xi'-\xi)&\quad
\mbox{if $(q,T,\xi)\prec (q',T',\xi')$,}\\
0                                &\quad\mbox{otherwise.}
\end{array}
\right.
\end{equation}

A configuration $(q,T,\xi)$ is said to be {\em locally like\/} 
a configuration $(q',T',\xi')$ if $q=q'$ and 
$T(\xi+d)=T'(\xi'+d)$ for all $d\in\{-1,0,1\}$.

\begin{Lemma}\label{th:825b}
For any $C_{1},C_{2}\in{\cal C}(Q,\Sigma)$, 
if they are locally like each other,
we have
$$
\bra{C_{1}}M_{\delta}M_{\delta}^{\dagger}\ket{C_{1}}=
\bra{C_{2}}M_{\delta}M_{\delta}^{\dagger}\ket{C_{2}}.
$$
\end{Lemma}\label{th:622b}
\begin{Proof}
Let $\tau_{-1},\tau_{0},\tau_{1}\in \Sigma$. 
Suppose that a configuration $C'=(p,T',\xi')$ is such that
$T'(\xi'-d)=\tau_{d}$ for all $d\in\{-1,0,1\}$.
Since every configuration locally like $C'$ also satisfies the above
condition, it suffices to show that
 $\bra{C'}M_{\delta}M_{\delta}^{\dagger}\ket{C'}$
depends only on $p,\tau_{-1},\tau_{0},\tau_{1}$.
By Proposition \ref{th:825a} and Eq.\ (\ref{eq:44}) we have
\begin{eqnarray*}
\bra{C'}M_{\delta}M_{\delta}^{\dagger}\ket{C'}
&=&
\sum_{C\in{\cal C}(Q,\Sigma)}|\bra{C'}M_{\delta}\ket{C}|^{2}\\
&=&
\sum_{C\prec C'}|\bra{C'}M_{\delta}\ket{C}|^{2}\\
&=&
\sum_{q,\sigma,d}
|\bra{C'}M_{\delta}\ket{\beta(q,\sigma,d)C'}|^{2}\\
&=&
\sum_{q,\sigma,d}
|\bra{p,T',\xi'}M_{\delta}\ket{q,{T'}^{\sigma}_{\xi'-d},\xi'-d}|^{2}\\
&=&
\sum_{q,\sigma,d}
|\delta(q,{T'}^{\sigma}_{\xi'-d}(\xi'-d),p,T'(\xi'-d),d)|^{2}\\
&=&
\sum_{q,\sigma,d}
|\delta(q,\sigma,p,\tau_{d},d)|^{2}.
\end{eqnarray*}
The first equality above follows from Parseval's identity. 

Thus, $\bra{C'}M_{\delta}M_{\delta}^{\dagger}\ket{C'}$
depends only on $p,\tau_{-1},\tau_{0},\tau_{1}$ and the proof is
completed.
\end{Proof}

For the case where the head is required to move,
 a proof of the following lemma appeared first in \cite{BV97}.
 The following proof not only covers the general case but also 
simplifies the argument given in \cite{BV97}. 

\begin{Lemma}\label{th:518a}
The evolution operator $M_{\delta}$ of a local transition function $\delta$
is unitary if it is an isometry.
\end{Lemma}

\begin{Proof}
Suppose that $M_{\delta}$ is an isometry,
 i.e., $M_{\delta}^{\dagger}M_{\delta}=1$.
Obviously, $M_{\delta}M_{\delta}^{\dagger}$ is a projection.
If $\bra{C}M_{\delta}M_{\delta}^{\dagger}\ket{C}=1$ for every 
$C\in{\cal C}(Q,\Sigma)$, the computational basis 
is included in the range of $M_{\delta}M_{\delta}^{\dagger}$
and then, since the range of any projection is a closed linear subspace, 
we have $M_{\delta}M_{\delta}^{\dagger}=1$ so that $M_{\delta}$ is unitary.
Thus, it suffices to show that 
$\bra{C}M_{\delta}M_{\delta}^{\dagger}\ket{C}=1$
 for every $C\in{\cal C}(Q,\Sigma)$.
To show this, suppose that there is a configuration
 $C_{0}\in{\cal C}(Q,\Sigma)$ 
such that $\bra{C_{0}}M_{\delta}M_{\delta}^{\dagger}\ket{C_{0}}
=1-\epsilon$ with $\epsilon>0$.
For any $n>2$ and $d\in\{-1,0,1\}$,
let $S(n,d)$ be the set of configurations such that
\begin{eqnarray*}
S(n,d)
=\{(q,T,\xi)\in{\cal C}(Q,\Sigma)|\ T(m)=B\!\!\!
&\mbox{for}&\!\!\!\!\mbox{all}\ \ m\not\in\{1,\ldots,n\}\\
&\mbox{and}&\!\!\!\xi\in\{1-d,\ldots,n+d\} \}.
\end{eqnarray*}
Let 
\begin{equation}\label{eq:45}
A=\sum_{(C,C')\in S(n,0)\times S(n,1)}|\bra{C'}M_{\delta}\ket{C}|^{2}
\end{equation}
and we shall consider evaluations of $A$ in terms of the numbers
of elements of the sets $S(n,0)$ and $S(n,1)$.
It is easy to see that if $C\in S(n,0)$ and $C\prec C'$ then
$C'\in S(n,1)$.
It follows from Eq.\ (\ref{eq:44}) 
that $\bra{C'}M_{\delta}\ket{C}=0$ for any pair
$(C,C')$ with $C\in S(n,0)$ and $C'\not\in S(n,1)$ so that 
the summation over $(C,C')\in S(n,0)\times S(n,1)$ in Eq.\ (\ref{eq:45}) 
can be replaced by the summation over
 $(C,C')\in S(n,0)\times {\cal C}(Q,\Sigma)$.
By Parseval's identity, we have
$$
A=\sum_{(C,C')\in S(n,0)\times {\cal C}(Q,\Sigma)}
|\bra{C'}M_{\delta}\ket{C}|^{2}
=\sum_{C\in S(n,0)}\bra{C}M_{\delta}^{\dagger}M_{\delta}{\ket{C}}.
$$
Since $M_{\delta}$ is an isometry, we have
$$
A=|S(n,0)|.
$$
Let $S(C_{0})$ be the set of all configurations in $S(n,-1)$
locally like $C_{0}$.  Then, $S(C_{0})\subseteq S(n,1)$.
By Lemma \ref{th:825b}, 
$\bra{C'}M_{\delta}M_{\delta}^{\dagger}\ket{C'}=1-\epsilon$
 for all $C'\in S(C_{0})$.
Thus, we have
\begin{eqnarray*}
A
&\le&\sum_{(C,C')\in {\cal C}(Q,\Sigma)\times S(n,1)}
|\bra{C'}M_{\delta}\ket{C}|^{2}\\
&=&\sum_{C'\in S(n,1)}\bra{C'}M_{\delta}M_{\delta}^{\dagger}\ket{C'}\\
&\le& (1-\epsilon)|S(C_{0})|+|S(n,1)|-|S(C_{0})|\\
&=&|S(n,1)|-\epsilon|S(C_{0})|.
\end{eqnarray*}
The cardinalities of $S(n,d)$ and $S(C_{0})$ are given by
$|S(n,d)|=(n+2d)|Q|\,|\Sigma|^{n}$ and
$|S(C_{0})|=(n-2)|\Sigma|^{n-3}$.
Therefore, we have
$$
|\Sigma|^{n-3}(2|Q||\Sigma|^{3}-\epsilon(n-2))
=|S(n,1)|-\epsilon|S(C_{0})|-|S(n,0)|
\ge 0
$$
for all $n>2$.
But, for $n>2+2\epsilon^{-1}|Q||\Sigma|^{3}$,
this yields an obvious contradiction and the proof is completed.
\end{Proof}

According to discussions in Section 3, a quantum Turing machine
can be defined as a mathematical structure $(Q,\Sigma,\delta)$ consisting of
a Turing frame $(Q,\Sigma)$ and a local transition function $\delta$
such that the evolution operator $M_{\delta}$ is unitary. 
The following theorem characterizes intrinsically the local transition 
functions that give rise to quantum Turing machines.

\begin{Theorem}\label{th:518b}
The evolution operator $M_{\delta}$ of a local transition function $\delta$
for the Turing frame $(Q,\Sigma)$ is unitary if and only if $\delta$ 
satisfies the following conditions.

{\rm (a)} For any $(q,\sigma)\in Q\times\Sigma$,
$$
\sum_{p,\tau,d}|\delta(q,\sigma,p,\tau,d)|^{2}=1.
$$

{\rm (b)} For any $(q,\sigma), (q',\sigma')\in Q\times\Sigma$
with $(q,\sigma)\ne (q',\sigma')$,
$$
\sum_{p,\tau,d}
\delta(q',\sigma',p,\tau,d)^{*}\delta(q,\sigma,p,\tau,d)=0.
$$

{\rm (c)} For any $(q,\sigma,\tau),(q',\sigma',\tau')\in Q\times\Sigma^{2}$,
 we have
$$
\sum_{p\in Q,d=0,1}\delta(q',\sigma',p,\tau',d-1)^{*}
\delta(q,\sigma,p,\tau,d)=0.
$$

{\rm (d)} For any $(q,\sigma,\tau),(q',\sigma',\tau')\in Q\times\Sigma^{2}$,
 we have
$$
\sum_{p\in Q}\delta(q',\sigma',p,\tau',-1)^{*}\delta(q,\sigma,p,\tau,1)=0.
$$
\end{Theorem}

\begin{Proof}
Let $\delta$ be a local transition function for a Turing frame $(Q,\Sigma)$.
Let $C=(q,T,\xi)\in {\cal C}(Q,\Sigma)$.
From Eq.\ (\ref{eq:43}) we have
\begin{eqnarray*}
\lefteqn{\bra{C}M_{\delta}^{\dagger}M_{\delta}\ket{C}}\quad\\
&=&
\sum_{p,\tau,d}\sum_{p',\tau',d'}
\delta(q,T(\xi),p',\tau',d')^{*}\delta(q,T(\xi),p,\tau,d)
\braket{p',T^{\tau'}_{\xi},\xi+d'|p,T^{\tau}_{\xi},\xi+d}\\
&=&
\sum_{p,\tau,d}|\delta(q,T(\xi),p,\tau,d)|^{2}.
\end{eqnarray*}
Since for any $\sigma\in\Sigma$
 there are some $T\in\Sigma^{\#}$ and $\xi\in\Z$
such that $T(\xi)=\sigma$, condition (a) holds
if and only if $\bra{C}M_{\delta}^{\dagger}M_{\delta}\ket{C}=1$ for any 
$C\in{\cal C}(Q,\Sigma)$.

Let $C=(q,T,\xi)\in{\cal C}(Q,\Sigma)$ and
 $C'=(q',T',\xi')\in{\cal C}(Q,\Sigma)$.
From Eq.\ (\ref{eq:43}) we have
\begin{eqnarray*}
\lefteqn{\bra{C'}M_{\delta}^{\dagger}M_{\delta}\ket{C}}\quad\\
&=&\!\!\!\!\!
\sum_{p,\tau,d}\sum_{p',\tau',d'}
\delta(q',T'(\xi'),p',\tau',d')^{*}\delta(q,T(\xi),p,\tau,d)
\braket{p',{T'}^{\tau'}_{\xi'},\xi'+d'|p,T^{\tau}_{\xi},\xi+d}\\
&=&\!\!\!\!\!
\sum{}^{*}
\delta(q',T'(\xi'),p,\tau',d')^{*}\delta(q,T(\xi),p,\tau,d),
\end{eqnarray*}
where the summation $\sum^{*}$ is taken over all $p\in Q$,
$\tau,\tau'\in\Sigma$, and $d,d'\in\{-1,0,1\}$ such that 
$T^{\tau}_{\xi}={T'}^{\tau'}_{\xi'}$ and $\xi+d=\xi'+d'$.

For any $k\in\Z$,
let ${\cal C}(k)$ be a subset of ${\cal C}(Q,\Sigma)^{2}$ consisting of all
pairs $C=(q,T,\xi)$ and $C'=(q',T',\xi')$ with $C\not=C'$ 
such that $T(m)=T'(m)$ for all $m\not\in\{\xi,\xi'\}$ and that 
$\xi'-\xi=k$.
It is easy to see that if $C\neq C'$ and
$$
(C,C')\not\in\bigcup_{k\in\{0,\pm 1,\pm 2\}}{\cal C}(k)
$$
then $\bra{C'}M_{\delta}^{\dagger}M_{\delta}\ket{C}=0$.
We shall show that condition (b), (c), or (d) holds if and only if
$\bra{C'}M_{\delta}^{\dagger}M_{\delta}\ket{C}=0$ holds
 for all $(C,C')\in{\cal C}(0)$,
$(C,C')\in{\cal C}(1)$, or $(C,C')\in{\cal C}(2)$, respectively.

For any $(C,C')\in{\cal C}(0)$ with $C=(q,T,\xi)$ and $C'=(q',T',\xi')$, 
we have $T^{\tau}_{\xi}={T'}^{\ \tau'}_{\xi'}$ and $\xi+d=\xi'+d'$
if and only if $\tau=\tau'$ and $d=d'$,
so that we have
\begin{eqnarray*}
\bra{C'}M_{\delta}^{\dagger}M_{\delta}\ket{C}
&=&\sum_{p,\tau,d}\delta(q',T'(\xi'),p,\tau,d)^{*}\delta(q,T(\xi),p,\tau,d).
\end{eqnarray*}
Since for any $(q,\sigma),(q',\sigma')\in Q\times \Sigma$ with 
$(q,\sigma)\not=(q',\sigma')$
there are configurations
$C=(q,T,\xi)$ and $C'=(q',T',\xi')$ such that
$(C,C')\in {\cal C}(0)$, $T(\xi)=\sigma$ and $T'(\xi')=\sigma'$, 
condition (b) holds
if and only if $\bra{C'}M_{\delta}^{\dagger}M_{\delta}\ket{C}=0$ for all 
$(C,C')\in{\cal C}(0)$.

For any $(C,C')\in{\cal C}(1)$ with $C=(q,T,\xi)$ and $C'=(q',T',\xi')$, 
we have $T^{\tau}_{\xi}={T'}^{\tau'}_{\xi'}$ and
$\xi+d=\xi'+d'$
if and only if $\tau=T'(\xi)$, $\tau'=T(\xi')$,
and  $(d,d')\in\{(0,-1),(1,0)\}$, so that we have
\begin{eqnarray*}
\bra{C'}M_{\delta}^{\dagger}M_{\delta}\ket{C}
&=&\sum_{p\in Q,d=0,1}\delta(q',T'(\xi'),p,T(\xi'),d-1)^{*}
\delta(q,T(\xi),p,T'(\xi),d).
\end{eqnarray*}
Since for any $(q,\sigma,\tau),(q',\sigma',\tau')\in Q\times \Sigma^{2}$ 
there are configurations $C=(q,T,\xi)$ and $C'=(q',T',\xi')$ such that
$C,C'\in{\cal C}(1)$,
$(T(\xi),T'(\xi))=(\sigma,\tau)$, and
$(T'(\xi'),T(\xi'))=(\sigma',\tau')$, 
condition (c) holds if and only if 
$\bra{C'}M_{\delta}^{\dagger}M_{\delta}\ket{C}=0$ for any 
$(C,C')\in{\cal C}(1)$.

For any $(C,C')\in{\cal C}(2)$ with $C=(q,T,\xi)$ and $C'=(q',T',\xi')$, 
we have $T^{\tau}_{\xi}={T'}^{\tau'}_{\xi'}$ and  
$\xi+d=\xi'+d'$
if and only if $\tau=T'(\xi)$, $\tau'=T(\xi')$, $d=1$, 
and $d'=-1$, so that we have
\begin{eqnarray*}
\bra{C'}M_{\delta}^{\dagger}M_{\delta}\ket{C}
&=&\sum_{p\in Q}\delta(q',T'(\xi'),p,T(\xi'),-1)^{*}
\delta(q,T(\xi),p,T'(\xi),1).
\end{eqnarray*}
Thus,
condition (d) holds if and only if 
$\bra{C'}M_{\delta}^{\dagger}M_{\delta}\ket{C}=0$ for all 
$(C,C')\in{\cal C}(2)$.

Since $M_{\delta}^{\dagger}M_{\delta}$ is self-adjoint, 
$M_{\delta}$ is an isometry if and only if $
\bra{C'}M_{\delta}^{\dagger}M_{\delta}\ket{C}$ $=$ $\langle C'|C\rangle$ 
for any $C=(q,T,\xi),\ C'=(q',T',\xi')\in{\cal C}(Q,\Sigma)$
 with $\xi\le \xi'$.
Therefore, we have proved that conditions (a)--(d) hold if and
only if $M_{\delta}$ is an isometry.
Now, Lemma \ref{th:518a} concludes the assertion.
\end{Proof}

A quantum Turing machine $M=(Q,\Sigma,\delta)$ is called {\em unidirectional},
if we have $d=d'$ 
whenever $\delta(q,\sigma,p,\tau,d)\delta(q',\sigma',p,\tau',d')
\not=0$ for any $q,q'\in Q$, $\sigma,\sigma', \tau,\tau'\in\Sigma$, 
and $d,d'\in \{-1,0,1\}$.
It is easy to see that conditions (c) and (d) are automatically satisfied
by every unidirectional quantum Turing machine.
Thus, if every quantum Turing machine can be efficiently simulated 
by a unidirectional one without error, complexity theoretical 
consideration on quantum Turing machines can be done much easier. 
For two-way quantum Turing machines, this was shown by 
Bernstein and Vazirani \cite{BV97}.  For general quantum Turing
machines defined by the above conditions,  the positive answer 
will be given in our forthcoming paper \cite{NO99}, 
including extension to multi-tape quantum Turing machines 
defined by the conditions of Theorem 6.2.

\section{Alternative approaches to the characterization of 
local transition functions}

Hirvensalo \cite{Hir97} gave 
the following set of conditions for a local transition function
$\delta$ to give the unitary evolution operator (see also \cite{Gru99}):

{\rm (H-a)} For any $(q,\sigma)\in Q\times\Sigma$,
$$
\sum_{p,\tau,d}|\delta(q,\sigma,p,\tau,d)|^{2}=1.
$$

{\rm (H-b)} For any $(q,\sigma), (q',\sigma')\in Q\times\Sigma$
with $(q,\sigma)\ne (q',\sigma')$,
$$
\sum_{p,\tau,d}
\delta(q',\sigma',p,\tau,d)^{*}\delta(q,\sigma,p,\tau,d)=0.
$$

{\rm (H-c)} For any $(p,\tau,d),(p',\tau',d')\in Q\times\Sigma\times\{-1,0,1\}$
 with $(p,\tau,d)\ne (p',\tau',d')$, we have
$$
\sum_{(q,\sigma)\in Q\times\Sigma}\delta(q,\sigma,p,\tau,d)^{*}
\delta(q,\sigma,p',\tau',d')=0.
$$

{\rm (H-d)} For any $(q,\sigma,\tau),(q',\sigma',\tau')\in Q\times\Sigma^{2}$
 and $d\neq d'\in\{-1,0,1\}$, we have
$$
\sum_{p\in Q}\delta(q,\sigma,p,\tau,d)^{*}\delta(q',\sigma',p,\tau',d')=0.
$$

However, the above set of conditions consists of only a sufficient 
condition, not a necessary one.  To show this, let $Q=\{0,1\}$, 
$\Sigma=\{B\}$, and define a local transition function $\delta$ as follows.

$$ 
\begin{array}{lll}
\delta(0,B,0,B,-1)=0,& \delta(0,B,0,B,0)=1/2,& \delta(0,B,0,B,1)=-1/2,\\
\delta(0,B,1,B,-1)=1/2,& \delta(0,B,1,B,0)=1/2,& \delta(0,B,1,B,1)=0,\\
\delta(1,B,0,B,-1)=0,& \delta(1,B,0,B,0)=1/2,& \delta(1,B,0,B,1)=1/2,\\
\delta(1,B,1,B,-1)=1/2,& \delta(1,B,1,B,0)=-1/2,& \delta(1,B,1,B,1)=0.
\end{array}
$$

Then, $\delta$ satisfies conditions (a)--(d) of Theorem \ref{th:518b} 
and hence gives the unitary evolution operator, but does not satisfy
Hirvensalo's conditions.  In fact, $\delta$ does not satisfy condition (H-c),
since 
$$
\delta(0,B,0,B,0)^{*}\delta(0,B,1,B,-1)
+\delta(1,B,0,B,0)^{*}\delta(1,B,1,B,-1)
=1/2,
$$
and $\delta$ does not satisfy condition (H-d), since
$$
\delta(0,B,0,B,0)^{*}\delta(1,B,0,B,1)
+\delta(0,B,1,B,0)^{*}\delta(1,B,1,B,1)=1/4.
$$
Thus, conditions (H-d) and (H-c) are not necessary.

The conditions in Theorem \ref{th:518a} are obtained from the requirement
that the column vectors of the evolution operator are orthonormal in
the matrix representation in the computational basis. 
Hirvensalo's conditions mix requirements for column vectors and 
for row vectors.  In the rest of this section,
 from the sole requirement that the row vectors are orthonormal,
 we shall obtain a set of necessary and 
sufficient conditions for the unitarity of the evolution operator.

The proof of the following lemma is similar to that of Lemma \ref{th:518a}.

\begin{Lemma}\label{th:4} 
The evolution operator $M_\delta$ of a local transition function $\delta$
 is unitary if $M_\delta M_\delta^{\dagger}=1$.
\end{Lemma}

Now we give another characterization of the local transition functions
 that give rise to quantum Turing machines.
\begin{Theorem}\label{th:5} 
 The evolution operator $M_\delta$ of a local transition function $\delta$
 for the Turing frame $(Q,\Sigma)$ is unitary 
if and only if $\delta$ satisfies the following conditions.

{\rm (a)} For any $p\in Q$ and $\tau_{-1},\tau_0,\tau_1\in\Sigma$,
$$
 \sum_{q\in Q,\sigma\in\Sigma,d\in\{-1,0,1\}}|\delta(q,\sigma,p,\tau_d,d)|^2=1.
$$

{\rm (b)} For any $p,p'\in Q$ with $p\neq p'$
 and any $\tau_{-1},\tau_0,\tau_1\in\Sigma$,
$$
 \sum_{q\in Q,\sigma\in\Sigma,d\in\{-1,0,1\}}\delta(q,\sigma,p',\tau_d,d)^*
\delta(q,\sigma,p,\tau_d,d)=0. 
$$

{\rm (c)} For any $p,p'\in Q$ and $\tau_0,\tau_1\in\Sigma$,
$$
 \sum_{q\in Q,\sigma\in\Sigma,d=0,1}\delta(q,\sigma,p',\tau_d,d-1)^*
\delta(q,\sigma,p,\tau_d,d)=0.
$$

{\rm (d)} For any $(p,\tau),(p',\tau')\in Q\times \Sigma$
 with $\tau\neq\tau'$ and any $d\in\{-1,0,1\}$, we have
$$
 \sum_{q\in Q,\sigma\in\Sigma}
\delta(q,\sigma,p',\tau',d)^{*}\delta(q,\sigma,p,\tau,d)=0.
$$

{\rm (e)} For any $(p,\tau),(p',\tau')\in Q\times \Sigma$
 with $\tau\neq\tau'$ and any $d=0,1$, we have
$$
 \sum_{q\in Q,\sigma\in\Sigma}
\delta(q,\sigma,p',\tau',d-1)^{*}\delta(q,\sigma,p,\tau,d)=0.
$$

{\rm (f)} For any $(p,\tau),(p',\tau')\in Q\times\Sigma$, we have
$$ 
\sum_{q\in Q,\sigma\in\Sigma}
\delta(q,\sigma,p',\tau',-1)^*\delta(q,\sigma,p,\tau,1)=0. 
$$
\end{Theorem}

\begin{Proof} Let $\delta$ be a local transition function for a Turing frame
 $(Q,\Sigma)$. Let $C=(p,T,\xi)\in{\cal C}(Q,\Sigma)$. 
From Proposition \ref{th:825a} and Eq.\ (\ref{eq:44}), we have
\begin{eqnarray}\label{eq:51}
 M_\delta^{\dagger}|p,T,\xi\rangle &=& 
\sum_{C'\in{\cal C}(Q,\Sigma)}|C'\rangle
\langle C'|M_\delta^{\dagger}|C\rangle\nonumber\\
&=& \sum_{C':C'\prec C}{\langle C|M_\delta|C'\rangle}^*|C'\rangle\nonumber\\
&=& \sum_{q,\sigma,d}{\langle C|M_\delta|\beta(q,\sigma,d)C\rangle}^*
\ket{\beta(q,\sigma,d)C}\nonumber\\
&=& \sum_{q,\sigma,d}
{\langle p,T,\xi|M_\delta|q,T_{\xi-d}^{\sigma},\xi-d\rangle}^*
\ket{q,T_{\xi-d}^{\sigma},\xi-d}\nonumber\\
&=& \sum_{q,\sigma,d}\delta(q,T_{\xi-d}^\sigma(\xi-d),p,T(\xi-d),d)^*
\ket{q,T_{\xi-d}^{\sigma},\xi-d}\nonumber\\
&=& \sum_{q,\sigma,d}\delta(q,\sigma,p,T(\xi-d),d)^*
\ket{q,T_{\xi-d}^\sigma,\xi-d}.
\end{eqnarray}

From Eq.\ (\ref{eq:51}) we have
\begin{eqnarray*}
\langle C|M_\delta M_\delta^{\dagger}|C\rangle
&=& \sum_{q,\sigma,d}\sum_{q',\sigma',d'}
\delta(q,\sigma,p,T(\xi-d),d)^*
\delta(q',\sigma',p,T(\xi-d'),d')\\
& & \times \langle q',T_{\xi-d'}^{\sigma'},\xi-d'
\ket{q,{T}_{\xi-d}^{\sigma},\xi-d}
\\
&=& \sum_{q,\sigma,d}|\delta(q,\sigma,p,T(\xi-d),d)|^2.
\end{eqnarray*}
Since for any $\tau_{-1},\tau_0,\tau_1\in\Sigma$ there are
 some $T\in\Sigma^\#$ and $\xi\in\Z$ such that $T(\xi-d)=\tau_d$,
 condition (a) holds if and only if
 $\langle C|M_\delta M_\delta^{\dagger}|C\rangle=1$
 for any $C\in{\cal C}(Q,\Sigma)$.
 
Let $C=(p,T,\xi)\in{\cal C}(Q,\Sigma)$ and
 $C'=(p',T',\xi')\in{\cal C}(Q,\Sigma)$.
 From Eq.\ (\ref{eq:51}) we have
\begin{eqnarray*}
\langle C|M_\delta M_\delta^{\dagger}|C'\rangle
&=& \sum_{q,\sigma,d}\sum_{q',\sigma',d'}
\delta(q',\sigma',p',T'(\xi'-d'),d')^*
\delta(q,\sigma,p,T(\xi-d),d)\\
& & \times
\langle q,T_{\xi-d}^{\sigma},\xi-d
|q',{T'}_{\xi'-d'}^{\sigma'},\xi'-d'\rangle
\\
&=& \sum{}^{**}\delta(q,\sigma,p',T'(\xi'-d'),d')^*
\delta(q,\sigma,p,T(\xi-d),d),
\end{eqnarray*}
where the summation $\sum^{**}$ is taken over all $q\in Q$, $\sigma\in\Sigma$,
 and $d,d'\in\{-1,0,1\}$
 such that $T_{\xi-d}^{\sigma}={T'}_{\xi'-d'}^{\sigma}$ and $\xi-d=\xi'-d'$.

For any $k\in\Z$ and $d\in\{-1,0,1\}$, let
 $A(k)$ be a subset of ${\cal C}(Q,\Sigma)^2$
 consisting of all pairs $C=(p,T,\xi)$
 and $C'=(p',T',\xi')$ with $C\neq C'$, $T=T'$ and $\xi-\xi'=k$, and
 $B(k,d)$ be a subset of
 ${\cal C}(Q,\Sigma)^2$ consisting of all pairs $C=(p,T,\xi)$ and
 $C'=(p',T',\xi')$ with $T\neq T'$ and $\xi-\xi'=k$
 such that $T(m)=T'(m)$ for all $m\neq\xi-d$.
 It is easy to see that if $C\neq C'$ and
$$ 
(C,C')\not\in\left(\bigcup_{k\in\{0,\pm 1,\pm 2\}}A(k)\right)\cup\left(
\bigcup_{(k,d):|k-d|\le 1}B(k,d)\right) 
$$
then $\langle C|M_\delta M_\delta^{\dagger}|C'\rangle=0$.
 Let $B(0)=B(0,1)\cup B(0,0)\cup B(0,-1)$ and $B(1)=B(1,1)\cup B(1,0)$.
 We shall show that condition (b), (c), (d), (e) or (f) holds
 if and only if $\langle C|M_\delta M_\delta^{\dagger}|C'\rangle=0$ holds
 for all $(C,C')\in A(0)$,\ $(C,C')\in A(1)$,\ $(C,C')\in B(0)$,\ 
 $(C,C')\in B(1)$, or $(C,C')\in A(2)\cup B(2,1)$,
 respectively.

For any $(C,C')\in A(0)$ with $C=(p,T,\xi)$ and $C'=(p',T,\xi)$,
 we have $T_{\xi-d}^\sigma={T}_{\xi-d'}^{\sigma}$ and $\xi-d=\xi-d'$
 if and only if $d=d'$, so that we have
$$ 
\langle C|M_\delta M_\delta^{\dagger}|C'\rangle=\sum_{q,\sigma,d}
\delta(q,\sigma,p',T(\xi-d),d)^*\delta(q,\sigma,p,T(\xi-d),d). 
$$
Since for any $p,p'\in Q$ with $p\neq p'$ and
 any $\tau_{-1},\tau_0,\tau_1\in\Sigma$
 there are configurations $C=(p,T,\xi)$ and $C'=(p',T,\xi)$
 such that $(C,C')\in A(0)$ and $T(\xi-d)=\tau_d$ for all $d\in\{-1,0,1\}$,
 condition (b) holds if and only if
 $\langle C|M_\delta M_\delta^{\dagger}|C'\rangle=0$ holds
 for all $(C,C')\in A(0)$.

For any $(C,C')\in A(1)$ with $C=(p,T,\xi)$ and $C'=(p',T,\xi')$,
 we have $T_{\xi-d}^{\sigma}={T}_{\xi'-d'}^{\sigma}$ and $\xi-d=\xi'-d'$
 if and only if $(d,d')\in\{(1,0),(0,-1)\}$, so that we have 
$$ 
\langle C|M_\delta M_\delta^{\dagger}|C'\rangle=
\sum_{q\in Q,\sigma\in\Sigma,d=0,1}\delta(q,\sigma,p',T(\xi-d),d-1)^*
\delta(q,\sigma,p,T(\xi-d),d). 
$$
Since for any $p,p'\in Q$ and $\tau_0,\tau_1\in\Sigma$
 there are configurations $C=(p,T,\xi)$ and $C'=(p',T,\xi')$
 such that $(C,C')\in A(1)$ and $T(\xi-d)=\tau_d$ for all $d\in\{0,1\}$,
 condition (c) holds if and only if
 $\langle C|M_\delta M_\delta^{\dagger}|C'\rangle=0$ holds
 for all $(C,C')\in A(1)$.

For any $(C,C')\in B(0,1)$ with $C=(p,T,\xi)$ and $C'=(p',T',\xi)$,
 we have $T_{\xi-d}^{\sigma}={T'}_{\xi-d'}^{\sigma}$ and $\xi-d=\xi-d'$
 if and only if $d=d'=1$, because $T(\xi-1)\neq T'(\xi-1)$
 and $T_{\xi-d}^\sigma(\xi-1)={T'}_{\xi-d'}^{\sigma}(\xi-1)$. Thus we have 
$$ 
\langle C|M_\delta M_\delta^{\dagger}|C'\rangle=
\sum_{q\in Q,\sigma\in\Sigma}\delta(q,\sigma,p',T'(\xi-1),1)^*
\delta(q,\sigma,p,T(\xi-1),1). 
$$
Since for any $(p,\tau),(p',\tau')\in Q\times\Sigma$ with $\tau\neq\tau'$
 there are configurations $C=(p,T,\xi)$ and $C'=(p',T',\xi')$
 such that $(C,C')\in B(0,1),T(\xi-1)=\tau$, and $T'(\xi-1)=\tau'$,
 the case $d=1$ of condition (d) holds if and only if
 $\langle C|M_\delta M_\delta^{\dagger}|C'\rangle=0$ holds
 for all $(C,C')\in B(0,1)$.
 Similarly we can show the case $d=0$ or $d=-1$ of condition (d) holds
 if and only if $\langle C|M_\delta M_\delta^{\dagger}|C'\rangle=0$ holds
 for all $(C,C')\in B(0,0)$ or $B(0,-1)$.
 Thus condition (d) holds
 if and only if $\langle C|M_\delta M_\delta^{\dagger}|C'\rangle=0$ holds
 for all $(C,C')\in B(0)$.

For any $(C,C')\in B(1,1)$ with $C=(p,T,\xi)$ and $C'=(p',T',\xi')$,
 we have $T_{\xi-d}^{\sigma}={T'}_{\xi'-d'}^{\sigma}$ and $\xi-d=\xi'-d'$
 if and only if $d=1$ and $d'=0$, 
 because $T(\xi-1)\neq T'(\xi-1)$ and $T_{\xi-d}^\sigma(\xi-1)
={T'}_{\xi'-d'}^{\sigma}(\xi-1)$. Thus we have 
$$ 
\langle C|M_\delta M_\delta^{\dagger}|C'\rangle=
\sum_{q\in Q,\sigma\in\Sigma}\delta(q,\sigma,p',T'(\xi-1),0)^*
\delta(q,\sigma,p,T(\xi-1),1). 
$$
Since for any $(p,\tau),(p',\tau')\in Q\times\Sigma$ with $\tau\neq\tau'$
 there are configurations $C=(p,T,\xi)$ and $C'=(p',T',\xi')$
 such that $(C,C')\in B(1,1),T(\xi-1)=\tau$, and $T'(\xi-1)=\tau'$,
 the case $d=1$ of condition (e) holds if and only if
 $\langle C|M_\delta M_\delta^{\dagger}|C'\rangle=0$ holds
 for all $(C,C')\in B(1,1)$.
 Similarly we can show the case $d=0$ of condition (e) holds
 if and only if $\langle C|M_\delta M_\delta^{\dagger}|C'\rangle=0$ holds
 for all $(C,C')\in B(1,0)$. Thus condition (e) holds
 if and only if $\langle C|M_\delta M_\delta^{\dagger}|C'\rangle=0$ holds
 for all $(C,C')\in B(1)$.

For any $(C,C')\in A(2)\cup B(2,1)$ with $C=(p,T,\xi)$ and $C'=(p',T',\xi')$,
 we have $T_{\xi-d}^{\sigma}={T'}_{\xi'-d'}^{\sigma}$ and $\xi-d=\xi'-d'$
 if and only if $d=1$ and $d'=-1$, so that we have 
$$ 
\langle C|M_\delta M_\delta^{\dagger}|C'\rangle=
\sum_{q\in Q,\sigma\in\Sigma}\delta(q,\sigma,p',T'(\xi-1),-1)^*
\delta(q,\sigma,p,T(\xi-1),1). 
$$
Since for any $(p,\tau),(p',\tau')\in Q\times\Sigma$
 there are configurations $C=(p,T,\xi)$ and $C'=(p',T',\xi')$
 such that $(C,C')\in A(2)\cup B(2,1),T(\xi-1)=\tau$, and $T'(\xi-1)=\tau'$,
 condition (f) holds if and only if
 $\langle C|M_\delta M_\delta^{\dagger}|C'\rangle=0$ holds
 for all $(C,C')\in A(2)\cup B(2,1)$.

Since $M_\delta M_\delta^{\dagger}$ is self-adjoint,
 $M_\delta M_\delta^{\dagger}=1$
 if and only if $\langle C|M_\delta M_\delta^{\dagger}|C'\rangle$
 $=$ $\langle C|C'\rangle$
 for any $C=(p,T,\xi),\ C'=(p',T',\xi')\in{\cal C}(Q,\Sigma)$
 with $\xi'\le\xi$.
 Therefore, we have proved that conditions (a)--(f) hold if and only if
 $M_\delta M_\delta^{\dagger}=1$.
 Now, Lemma \ref{th:4} concludes the assertion.
\end{Proof}  

\section{Multi-tape quantum Turing machines}

In the preceding sections, we have discussed solely single tape 
quantum Turing machines, but our arguments can be adapted easily to
multi-tape quantum Turing machines, which are quantum analogues of 
multi-tape deterministic Turing machines.

First, we explain how to adapt our arguments to
 multi-tape quantum Turing machines
 by considering two-tape quantum Turing machines.
 A two-tape quantum Turing machine is a quantum system
 consisting of a processor, two bilateral infinite tapes with
 heads to read and write symbols on their tapes.
 In order to discuss local transition functions,
 we adapt the formal definitions as follows.
 Let $(Q,\Sigma_{1},\Sigma_{2})$ be a triple, called a
 {\em {two-tape} Turing frame}, 
consisting of a finite sets $Q$, $\Sigma_{1}$, and  $\Sigma_{2}$ 
with specific elements $B_{1}\in\Sigma_{1}$ and $B_{2}\in\Sigma_{2}$.
The {\em configuration space} of $(Q,\Sigma_{1},\Sigma_{2})$ is the product
set 
${\cal C}(Q,\Sigma_{1},\Sigma_{2})
=Q\times\Sigma_{1}^{\#}\times\Sigma_{2}^{\#}\times\Z^{2}$.
Thus, the configuration of a two-tape quantum Turing machine ${\cal Q}$ with
the frame $(Q,\Sigma_{1},\Sigma_{2})$ is determined by the processor 
configuration $q\in Q$, 
the first and second tape configurations $T_{1}\in\Sigma_{1}^{\#}$, 
$T_{2}\in\Sigma_{2}^{\#}$,
and the head positions $\xi_{1}\in\Z$, $\xi_{2}\in\Z$ in
the first and second tapes.
The {\em quantum state space} of $(Q,\Sigma_{1},\Sigma_{2})$ is the 
Hilbert space ${\cal H}(Q,\Sigma_{1},\Sigma_{2})$ generated by 
${\cal C}(Q,\Sigma_{1},\Sigma_{2})$.  
A {\em local transition function} for $(Q,\Sigma_{1},\Sigma_{2})$ is 
defined to be a complex-valued function on 
$Q\times\Sigma\times Q\times\Sigma\times\{-1,0,1\}^{2}$,
where $\Sigma=\Sigma_{1}\times\Sigma_{2}$.
The relation
 $\delta(q,(\sigma_1,\sigma_2),p,(\tau_1,\tau_2),(d_1,d_2))=c$ 
 can be interpreted as the following operation of ${\cal Q}$:
 if the processor is in the configuration $q$ 
and if the head of the $i$-th tape
 ($i=1,2$) reads the symbol $\sigma_i$, then it follows with the amplitude $c$
 that the processor configuration turns to $p$, the head of the $i$-th tape
 writes the symbol $\tau_i$ and moves one cell to the right if
 $d_i=1$, to the left if $d_i=-1$, or does not move if $d_i=0$.
 The {\em evolution operator} of $\delta$ is a linear operator
 $M_\delta$ on ${\cal H}(Q,\Sigma_{1},\Sigma_{2})$ such that
\begin{eqnarray*} 
\lefteqn{M_\delta|q,(T_1,T_2),(\xi_1,\xi_2)\rangle}\quad\\
&=&\!\!\!\!\!\sum
\delta(q,(T_1(\xi_1),T_2(\xi_2)),p,(\tau_1,\tau_2),(d_1,d_2))
|p,({T_1}_{\xi_1}^{\tau_1},{T_2}_{\xi_2}^{\tau_2}),(\xi_1+d_1,\xi_2+d_2)\rangle
\end{eqnarray*} 
 for all $(q,(T_1,T_2),(\xi_1,\xi_2))\in {\cal C}(Q,\Sigma_{1},\Sigma_{2})$,
 where the summation is taken over all
 $(p,(\tau_1,\tau_2),(d_1,d_2))\in Q\times\Sigma\times\{-1,0,1\}^2$.
 Then, local transition functions of two-tape quantum Turing 
machines are characterized as follows.

\begin{Theorem}\label{th:1112a}
The evolution operator $M_\delta$ of a local transition function
 $\delta$ for the two-tape Turing frame $(Q,\Sigma_{1},\Sigma_{2})$ is unitary 
if and only if $\delta$ satisfies the following conditions.
 
{\rm (1)} For any $(q,\sigma)\in Q\times\Sigma$,
$$ 
\sum_{p\in Q,\tau\in\Sigma,d_1,d_2\in \{-1,0,1\}}
|\delta(q,\sigma,p,\tau,(d_1,d_2))|^2=1. 
$$

{\rm (2)} For any $(q,\sigma),(q',\sigma')\in Q\times \Sigma$
 with $(q,\sigma)\neq(q',\sigma')$,
$$ 
\sum_{p\in Q,\tau\in\Sigma,d_1,d_2\in \{-1,0,1\}}
\delta(q',\sigma',p,\tau,(d_1,d_2))^{*}
\delta(q,\sigma,p,\tau,(d_1,d_2))=0. 
$$

{\rm (3)} For any $(q,\sigma,\tau_2),(q',\sigma',\tau'_2)
\in Q\times\Sigma\times\Sigma_2$,
$$ 
\sum_{
\begin{array}{c}
{\scriptstyle p\in Q,\tau_1\in\Sigma_1}\\
{\scriptstyle d_1\in\{-1,0,1\},d_2=0,1}
\end{array}
}
\delta(q',\sigma',p,(\tau_1,\tau'_2),(d_1,d_2-1))^*
\delta(q,\sigma,p,(\tau_1,\tau_2),(d_1,d_2))=0. 
$$

{\rm (4)} For any $(q,\sigma,\tau_2),(q',\sigma',\tau'_2)
\in Q\times\Sigma\times\Sigma_2$,
$$ 
\sum_{p\in Q,\tau_1\in\Sigma_1,d_1\in\{-1,0,1\}}
\delta(q',\sigma',p,(\tau_1,\tau'_2),(d_1,-1))^*
\delta(q,\sigma,p,(\tau_1,\tau'_2),(d_1,1))=0. 
$$

{\rm (5)} For any $(q,\sigma,\tau),(q',\sigma',\tau')\in Q\times \Sigma^2$,
$$
\sum_{p\in Q,d_1=0,1}\delta(q',\sigma',p,\tau',(d_1-1,1))^*
\delta(q,\sigma,p,\tau,(d_1,-1))= 0.
$$

{\rm (6)} For any $(q,\sigma,\tau),(q',\sigma',\tau')\in Q\times \Sigma^2$,
$$
 \sum_{p\in Q,d_1=0,1,d_2=0,1}
\delta(q',\sigma',p,\tau',(d_1-1,d_2))^*\delta(q,\sigma,p,\tau,(d_1,d_2-1))=0.
$$

{\rm (7)} For any $(q,\sigma,\tau_1),(q',\sigma',\tau'_1)
\in Q\times\Sigma\times\Sigma_1$,
$$ 
\sum_{
\begin{array}{c}
{\scriptstyle p\in Q,\tau_1\in\Sigma_1}\\
{\scriptstyle d_1=0,1,d_2\in\{-1,0,1\}}
\end{array}
}
\delta(q',\sigma',p,(\tau'_1,\tau_2),(d_1-1,d_2))^*
\delta(q,\sigma,p,(\tau_1,\tau_2),(d_1,d_2))=0. 
$$

{\rm (8)} For any $(q,\sigma,\tau),(q',\sigma',\tau')\in Q\times \Sigma^2$,
$$
 \sum_{p\in Q, d_1=0,1,d_2=0,1}
\delta(q',\sigma',p,\tau',(d_1-1,d_2-1))^*\delta(q,\sigma,p,\tau,(d_1,d_2))=0.
$$

{\rm (9)} For any $(q,\sigma,\tau),(q',\sigma',\tau')\in Q\times \Sigma^2$,
$$
\sum_{p\in Q,d_1=0,1}\delta(q',\sigma',p,\tau',(d_1-1,-1))^*
\delta(q,\sigma,p,\tau,(d_1,1))=0.
$$

{\rm (10)} For any $(q,\sigma,\tau),(q',\sigma',\tau')\in Q\times \Sigma^2$,
$$
 \sum_{p\in Q}\delta(q',\sigma',p,\tau',(-1,1))^*
\delta(q,\sigma,p,\tau,(1,-1))=0.
$$

{\rm (11)} For any $(q,\sigma,\tau),(q',\sigma',\tau')\in Q\times \Sigma^2$,
$$
\sum_{p\in Q,d_2=0,1}\delta(q',\sigma',p,\tau',(-1,d_2))^*
\delta(q,\sigma,p,\tau,(1,d_2-1))=0.
$$

{\rm (12)} For any $(q,\sigma,\tau_1),(q',\sigma',\tau'_1)
\in Q\times\Sigma\times\Sigma_1$,
$$ 
\sum_{p\in Q,\tau_2\in\Sigma_2,d_2\in\{-1,0,1\}}
\delta(q',\sigma',p,(\tau'_1,\tau_2),(-1,d_2))^*
\delta(q,\sigma,p,(\tau_1,\tau_2),(1,d_2))=0. 
$$

{\rm (13)} For any $(q,\sigma,\tau),(q',\sigma',\tau')\in Q\times \Sigma^2$,
$$
\sum_{p\in Q,d_2=0,1}\delta(q',\sigma',p,\tau',(-1,d_2-1))^*
\delta(q,\sigma,p,\tau,(1,d_2))= 0.
$$

{\rm (14)} For any $(q,\sigma,\tau),(q',\sigma',\tau')\in Q\times \Sigma^2$,
$$
\sum_{p\in Q}\delta(q',\sigma',p,\tau',(-1,-1))^*
\delta(q,\sigma,p,\tau,(1,1))= 0.
$$
\end{Theorem}

If each head is required to move either to the right
 or to the left at each step, conditions (3),(5)--(9),(11), and (13)
 are automatically satisfied. 
It is also easy to see that conditions (3)--(14) are automatically
satisfied by unidirectional two-tape quantum Turing machines,
for which $(d_{1},d_{2})$ is uniquely determined by $p$ in the 
non-zero amplitude $\delta(q,\sigma,p,\tau,(d_{1},d_{2}))$. 

The proof of Theorem \ref{th:1112a} is analogous to the proof of 
Theorem \ref{th:518b}. 
 Let ${\cal C}(k_1,k_2)$ be a subset of ${\cal C}(Q,\Sigma_{1},\Sigma_{2})^2$ 
consisting of all pairs 
$C=(q,(T_1,T_2),(\xi_1,\xi_2))$ and $C'=(q',(T'_1,T'_2),(\xi'_1,\xi'_2))$
 with $C\neq C'$ such that $T_{i}(m_i)=T'_{i}(m_i)$ 
for $m_{i}\not\in\{\xi_{i},\xi'_{i}\}$
 and that $\xi'_i-\xi_i=k_i$ for $i=1,2$.
 This plays a role similar to ${\cal C}(k)$
 in the proof of Theorem \ref{th:518b}.
 In the proof of Theorem \ref{th:518b},
 we showed that condition (b),(c), or (d) holds 
 if and only if $\langle C'|M_\delta^{\dagger}M_\delta|C\rangle=0$ holds
 for all $(C,C')\in{\cal C}(0),(C,C')\in{\cal C}(1),$
 or $(C,C')\in{\cal C}(2)$, respectively.
 In the case of Theorem \ref{th:1112a}, we can show similarly that
 for $(k_1,k_2)\in(\{0\}\times\{0,1,2\})
\cup(\{1,2\}\times\{0,\pm 1,\pm 2\})$,
 condition $(5k_{1}+k_{2}+2)$ holds if and only if 
 $\langle C'|M_\delta^{\dagger}M_\delta|C\rangle=0$ holds for all
 $(C,C')\in{\cal C}(k_1,k_2)$.
 For example, condition (2) holds if and only if 
 $\langle C'|M_\delta^{\dagger}M_\delta|C\rangle=0$ holds for all
 $(C,C')\in{\cal C}(0,0)$ (This is the case of $k_1=k_2=0$). 
 Moreover, it is trivial that 
 condition (1) holds if and only if
 $\langle C|M_\delta^{\dagger}M_\delta|C\rangle=1$ holds
 for all $C\in{\cal C}(Q,\Sigma_{1},\Sigma_{2})$,
 and that if $C\ne C'$ and
$$ 
(C,C')\not\in \bigcup_{(k_1,k_2)\in\{0,\pm 1,\pm 2\}^2}{\cal C}(k_1,k_2) 
$$
 then $\langle C'|M_\delta^{\dagger}M_\delta|C\rangle=0$. 
Since $M_\delta^{\dagger}M_\delta$ is self-adjoint, $M_\delta$ is an isometry 
if and only if $\langle C'|M_\delta^{\dagger}M_\delta|C\rangle
=\langle C'|C\rangle$ 
for any $C=(q,(T_1,T_2),(\xi_1,\xi_2))$, $C'=(q,(T'_1,T'_2),(\xi'_1,\xi'_2))
\in {\cal C}(Q,\Sigma_{1},\Sigma_{2})$ with $\xi_1<\xi'_1$
 or with $\xi_1=\xi'_1$ and
 $\xi_2\le\xi'_2$. Therefore, we can show that 
 conditions (1)--(14) hold if and only if $M_\delta$ is an isometry.
 We can also show that $M_\delta$ is unitary if it is an isometry 
by a similar argument with the proof of Lemma \ref{th:518a}.
 Thus we can prove Theorem \ref{th:1112a}.

 We now consider $k$-tape quantum Turing machines.
 In what follows, $\vec{a}$ abbreviates $(a_1,\ldots,a_k)$.
 For $j\in\{0,\ldots,k-1\}$, let $\vec{a}_{\le j}=(a_1,\ldots,a_j)$
 and $\vec{a}_{> j}=(a_{j+1},\ldots,a_k)$.
 For any set $S=\{i_1,\ldots,i_m\}\subseteq \{1,\ldots,k\}$, let
 $\vec{a}[S]=(a_{i_1},\ldots,a_{i_m})$ and
 $\bar{S}=\{1,\ldots,k\}\backslash S$. Moreover,
 for any tuple $(a_{i_1},\ldots,a_{i_m})$,
 the symbol $(a_{i_1},\ldots,a_{i_m})^t$
 denotes $(a_{I(1)},\ldots,a_{I(m)})$,
 where $\{I(1),\ldots,I(m)\}=\{i_1,\ldots,i_m\}$
 and $I(1)<\cdots<I(m)$.
 Extending the arguments for the two-tape quantum Turing machines,
 the local transition functions of $k$-tape quantum Turing machines
 can be characterized as follows.

\begin{Theorem}\label{th:009}
The evolution operator $M_\delta$ of a local transition function
 $\delta$ for the $k$-tape Turing frame
 $(Q,\Sigma_{1},\Sigma_{2},\ldots,\Sigma_{k})$ is unitary 
 if and only if $\delta$ satisfies the following conditions.

{\rm (1)} For any $(q,\vec{\sigma})\in Q\times\Sigma$,
$$ 
\sum_{p\in Q,\vec{\tau}\in\Sigma,\vec{d}\in \{-1,0,1\}^k}
|\delta(q,\vec{\sigma},p,\vec{\tau},\vec{d})|^2=1. 
$$

{\rm (2)} For any $(q,\vec{\sigma}),(q',\vec{\sigma'})\in Q\times \Sigma$
 with $(q,\vec{\sigma})\neq(q',\vec{\sigma'})$,
$$ 
\sum_{p\in Q,\vec{\tau}\in\Sigma,\vec{d}\in \{-1,0,1\}^k}
\delta(q',\vec{\sigma'},p,\vec{\tau},\vec{d})^*
\delta(q,\vec{\sigma},p,\vec{\tau},\vec{d})=0. 
$$

{\rm (3)} For each $j\in\{1,\ldots,k\}$ and
 $\vec{D}_{>k-j}=(D_{k-j+1},\ldots,D_k)
\in\{1,2\}\times\{0,\pm 1,\pm 2\}^{j-1}$,
 the following condition holds.
 For any $(q,\vec{\sigma},\vec{\tau}[S(\vec{D}_{>k-j})]),
(q',\vec{\sigma'},\vec{\tau'}[S(\vec{D}_{>k-j})])\in
 Q\times\Sigma\times \prod_{i\in S(\vec{D}_{>k-j})}\Sigma_i$ we have
\begin{eqnarray*}
\sum& &\!\!\!\!\!\delta(q',\vec{\sigma'},p,
(\vec{\tau}[\bar{S}(\vec{D}_{>k-j})],
\vec{\tau'}[S(\vec{D}_{>k-j})])^t,
(\vec{d}_{\le k-j},\vec{d'}_{>k-j}))^*\\
     & &\times\delta(q,\vec{\sigma},p,
(\vec{\tau}[\bar{S}(\vec{D}_{>k-j})],
\vec{\tau}[S(\vec{D}_{>k-j})])^t,
(\vec{d}_{\le k-j},\vec{d}_{>k-j}))=0,
\end{eqnarray*}
where the summantion is taken over $p\in Q$,
 $\vec{\tau}[\bar{S}(\vec{D}_{>k-j})]
\in\prod_{i\in \bar{S}(\vec{D}_{>k-j})}\Sigma_i$,
 $\vec{d}_{\le k-j}\in\{-1,0,1\}^{k-j}$,
 and $\vec{d'}_{>k-j},\vec{d}_{>k-j}\in\{-1,0,1\}^j$
 such that $\vec{d}_{>k-j}-\vec{d'}_{>k-j}=\vec{D}_{>k-j}$. 
 Here, $S(\vec{D}_{>k-j})=\{i\in\{k-j+1,\ldots,k\}|\ D_i\neq 0\}$.
\end{Theorem}

Note that condition (3) of Theorem \ref{th:009}
 contains $2\times\sum_{j=0}^{k-1}5^j$ conditions
 (the number of different pairs $(j,\vec{D}_{>k-j})$).
 Thus, the local transition functions of
 $k$-tape quantum Turing machines can be characterized by
$$ 
1+1+2\times\sum_{j=0}^{k-1}5^j=1+(1/2)(5^k+1)
$$ 
conditions.

Multi-tape Turing machines are often used for theoretical 
consideration in complexity theory \cite{Pap94}
because it is often easier to construct
 a multi-tape machine than a single tape machine
in order to realize a given algorithm.
Hence, multi-tape quantum Turing machines can be expected as useful 
tools for quantum complexity theory.
In such applications, it appears to be a tedious task to check that a
constructed local transition function satisfies the unitarity
conditions.  However, it should be noted that
 restricted classes of multi-tape machines
 are characterized much more simply; the unidirectional
multi-tape machines are characterized by only two conditions,
 conditions (1) and (2) in Theorem \ref{th:009}.

\appendix
\section{The bound of $M_\delta$}
\begin{Theorem} 
Let $\delta$ be a complex valued function defined on 
$Q\times\Sigma\times Q\times\Sigma\times\{-1,0,1\}$. 
Then, there is uniquely a bounded operator $M_\delta$ 
on ${\cal H}(Q,\Sigma)$ satisfying Eq.\ (\ref{eq:43}). 
The operator norm of $M_\delta$ is bounded by 
$\sqrt{5}K|Q||\Sigma|^{2}$, 
where
$$
K=\begin{array}{c}
 \\
{\rm max}\\
{}_{(q,\sigma)\in Q\times\Sigma}
\end{array}
\left(
\sum_{(p,\tau,d)\in Q\times\Sigma\times\{-1,0,1\}}
 |\delta(q,\sigma,p,\tau,d)|^2
\right)^{\frac{1}{2}}.
$$
\end{Theorem}

\begin{Proof}
For any $C=(q,T,\xi)\in {\cal C}(Q,\Sigma)$, 
let $\ket{F(C)}$ be defined by 
$$
\ket{F(C)}=
\sum_{p,\tau,d}\delta(q,T(\xi),p,\tau,d)\ket{p,T_\xi^\tau,\xi+d},
$$
where $(p,\tau,d)$ varies over 
the finite set $Q\times\Sigma\times\{-1,0,1\}$. 
Then we have 
$$
||\,\ket{F(C)}||^2 = \sum_{p,\tau,d} |\delta(q,T(\xi),p,\tau,d)|^2
                 \le K^2.
$$
By the Schwarz inequality, we have
$$
|\langle F(C')|F(C)\rangle|\le ||\,\ket{F(C')}||\cdot ||\,\ket{F(C)}||
                           \le K^2.
$$
for any $C,C'\in{\cal C}(Q,\Sigma)$. 
For any $(p,\tau)\in Q\times\Sigma$, 
let $\gamma_0(p,\tau)$ 
be the mapping on ${\cal C}(Q,\Sigma)$ defined by
$$
\gamma_0(p,\tau)(q,T,\xi)=(p,T_\xi^\tau,\xi). 
$$
According to Eq.\ (\ref{eq:41}), we have $\gamma_0(p,\tau)=\alpha(p,\tau,0)$. 
By Proposition 4.1 (iii), $\gamma_0(p,\tau)$ is 
a bijection between ${\cal C}(q,\sigma,0)$ and ${\cal C}(p,\tau,0)$. 
Thus, the operator 
$$
A_0(p,\tau;q,\sigma)=\sum_{C\in{\cal C}(q,\sigma,0)}
\langle F(\gamma_0(p,\tau)C)|F(C)\rangle 
\ket{\gamma_0(p,\tau)C}\bra{C}
$$ 
is bounded and its operator norm is at most $K^2$. 
By Proposition 4.1 (i),
$$
\sum_{(q,\sigma)\in Q\times\Sigma}
A_0(p,\tau;q,\sigma)
=
\sum_{C\in{\cal C}(Q,\Sigma)}
\langle F(\gamma_0(p,\tau)C)|F(C)\rangle 
\ket{\gamma_0(p,\tau)C}\bra{C},
$$
so that the operator
$$
A_0(p,\tau)
=
\sum_{C\in{\cal C}(Q,\Sigma)}
\langle F(\gamma_0(p,\tau)C)|F(C)\rangle 
\ket{\gamma_0(p,\tau)C}\bra{C}
$$
is bounded and its operator norm is at most $K^2|Q||\Sigma|$. 
Moreover, 
$$
A_0=\sum_{(p,\tau)\in Q\times\Sigma}A_0(p,\tau)
$$ 
is also bounded and $||A_0||\le K^2|Q|^2|\Sigma|^2$.

For any $(p,\tau,\tau')\in Q\times\Sigma^2$ and $i=\pm 1,\pm 2$, 
let $\gamma_i(p,\tau,\tau')$ be the mapping on ${\cal C}(Q,\Sigma)$ 
defined by
$$
\gamma_i(p,\tau,\tau')(q,T,\xi)=(p,T_{\xi,\xi+i}^{\tau,\tau'},\xi+i), 
%
%
$$
where $T_{\xi,\xi+i}^{\tau,\tau'}$ is the tape configuration defined by 
$$
T_{\xi,\xi+i}^{\tau,\tau'}(m)=\left\{
\begin{array}{ll}
\tau& \mbox{if}\ m=\xi,\\
\tau'& \mbox{if}\ m=\xi+i,\\
T(m)& \mbox{if}\ m\not=\xi,\xi+i.
\end{array}
\right.
$$
For any $(q,\sigma,\sigma')\in Q\times\Sigma^2$ and $i=\pm 1,\pm 2$, 
let ${\cal C}(q,\sigma,\sigma',i)$ be the set
$$
{\cal C}(q,\sigma,\sigma',i)
=\{ (q,T,\xi)\in {\cal C}(Q,\Sigma)|\ T(\xi)=\sigma
\ \mbox{and}\ T(\xi+i)=\sigma'\}.
$$
It is straightforward to check that 
$\gamma_i(p,\tau,\tau')$ is a bijection 
between ${\cal C}(q,\sigma,\sigma',i)$ and ${\cal C}(p,\tau',\tau,-i)$. 
Thus, for each $i=\pm 1,\pm 2$, the operator
$$
A_i(p,\tau,\tau';q,\sigma,\sigma')=\sum_{C\in{\cal C}(q,\sigma,\sigma',i)}
\langle F(\gamma_i(p,\tau,\tau')C)|F(C)\rangle 
\ket{\gamma_i(p,\tau,\tau')C}\bra{C}
$$ 
is bounded and $||A_i(p,\tau,\tau';q,\sigma,\sigma')||\le K^2$. 
For any $i\in\{\pm 1,\pm 2\}$, we can verify easily that 
if $(q,\sigma_1,\sigma_2)\neq (q',\sigma'_1,\sigma'_2)\in Q\times\Sigma^2$,
 then ${\cal C}(q,\sigma_1,\sigma_2,i)\cap {\cal C}(q',\sigma'_1,\sigma'_2,i)
=\emptyset$ and 
$$ 
{\cal C}(Q,\Sigma)=
\bigcup_{(q,\sigma,\sigma')\in Q\times\Sigma^2}{\cal C}(q,\sigma,\sigma',i).
$$
Thus, we have
$$
\sum_{(q,\sigma,\sigma')\in Q\times\Sigma^2}
A_i(p,\tau,\tau';q,\sigma,\sigma')
=
\sum_{C\in{\cal C}(Q,\Sigma)}
\langle F(\gamma_i(p,\tau,\tau')C)|F(C)\rangle 
\ket{\gamma_i(p,\tau,\tau')C}\bra{C},
$$
and the operator
$$
A_i(p,\tau,\tau')=\sum_{C\in{\cal C}(Q,\Sigma)}
\langle F(\gamma_i(p,\tau,\tau')C)|F(C)\rangle 
\ket{\gamma_i(p,\tau,\tau')C}\bra{C}
$$ 
is bounded and its operator norm is at most $K^2|Q||\Sigma|^2$. 
Moreover, 
$$
A_i=\sum_{(p,\tau,\tau')\in Q\times\Sigma^2}A_i(p,\tau,\tau')
$$ 
is also bounded and $||A_i||\le K^2|Q|^2|\Sigma|^4$.

Now, for $i=0,\pm 1,\pm 2$, let
$$ 
S(C,i)=\{(q',T',\xi')\in{\cal C}(Q,\Sigma)|\ \xi'-\xi=i\ \mbox{and}\ T(m)=T'(m)
\ \mbox{for any}\ m\not\in\{\xi,\xi'\}\},
$$ 
where $C=(q,T,\xi)\in {\cal C}(Q,\Sigma)$. Let
$$
A=\sum_{i=-2}^2\,\sum_{C'\in S(C,i)}\,\sum_{C\in {\cal C}(Q,\Sigma)}
\langle F(C')|F(C)\rangle \ket{C'}\bra{C}.
$$ 
Since for any $C'\in S(C,0)$ there is uniquely 
a pair $(p,\tau)\in Q\times\Sigma$ such that 
$C'=\gamma_0(p,\tau)C$ and for any $C'\in S(C,i)$, where $i=\pm 1,\pm 2$, 
there is uniquely a triple $(p,\tau,\tau')\in Q\times\Sigma^2$ 
such that $C'=\gamma_i(p,\tau,\tau')C$, we have 
\begin{eqnarray*}
A &=& \sum_{(p,\tau)\in Q\times\Sigma}\,\sum_{C\in{\cal C}(Q,\Sigma)}
\langle F(\gamma_0(p,\tau)C)|F(C)\rangle 
\ket{\gamma_0(p,\tau)C}\bra{C}\\
&+& \sum_{i\in\{\pm 1,\pm 2\}}\,\sum_{(p,\tau,\tau')\in Q\times\Sigma^2}
\,\sum_{C\in{\cal C}(Q,\Sigma)}
\langle F(\gamma_i(p,\tau,\tau')C)|F(C)\rangle 
\ket{\gamma_i(p,\tau,\tau')C}\bra{C}\\
&=& A_0+A_{1}+A_{-1}+A_2+A_{-2}
\end{eqnarray*}
is bounded and we have 
$$
||A||\le K^2|Q|^2|\Sigma|^2+4K^2|Q|^2|\Sigma|^4\le 5K^2|Q|^2|\Sigma|^4.
$$
Moreover, if $C'\not\in \bigcup_{i=0,\pm 1,\pm 2} S(C,i)$, 
then $\langle F(C')|F(C)\rangle=0$. Thus,
$$
A=\sum_{C',C\in {\cal C}(Q,\Sigma)}
\langle F(C')|F(C)\rangle \ket{C'}\bra{C}.
$$ 
For any $\ket{\psi}\in{\cal H}(Q,\Sigma)$, we have
\begin{eqnarray*}
\left|\left|
\sum_{C\in{\cal C}(Q,\Sigma)}
\langle C|\psi\rangle \ket{F(C)}
\right|\right|^2
&=&
\sum_{C',C\in{\cal C}(Q,\Sigma)}
\langle \psi|C'\rangle \langle C|\psi\rangle
\langle F(C)| F(C')\rangle \\
&=& 
\bra{\psi} A\ket{\psi}\\
&\le& 
5K^2|Q|^2|\Sigma|^4||\,\ket{\psi}||^2<\infty. 
\end{eqnarray*}
Now, let $M_\delta$ be an operator on ${\cal H}(Q,\Sigma)$ 
which transforms $\ket{\psi}$ to $\sum_{C\in{\cal C}(Q,\Sigma)}
\langle C|\psi\rangle \ket{F(C)}$. 
Then, $M_\delta$ is a unique bounded operator satisfying Eq.\ (\ref{eq:43}) 
and 
$$ 
||M_\delta||\le \sqrt{5}K|Q||\Sigma|^{2}.
$$
\end{Proof}

\end{document}